\RequirePackage{fix-cm}
\documentclass[11pt,twocolumn,english]{scrartcl}
\usepackage{mathpazo}

\usepackage[T1]{fontenc}
\usepackage[latin9]{inputenc}
\usepackage[letterpaper]{geometry}
\usepackage{units}
\usepackage{amsmath}
\usepackage{graphicx}

\makeatletter
\newcommand{\lyxaddress}[1]{
\par {\raggedright #1
\vspace{1.4em}
\noindent\par}
}

\@ifundefined{date}{}{\date{}}
\usepackage{numprint}
\usepackage[super]{nth}
\usepackage[superscript,biblabel]{cite}
\usepackage[font={sf,small}, labelfont={bf}, indention=0cm, format=plain]{caption}

\let\originalleft\left
\let\originalright\right
\renewcommand{\left}{\mathopen{}\mathclose\bgroup\originalleft}
\renewcommand{\right}{\aftergroup\egroup\originalright}

\let\Im\relax

\DeclareMathOperator\Im{Im}

\renewcommand\vec[1]{\ensuremath{\boldsymbol{#1}}}

\makeatother

\usepackage{babel}
\begin{document}

\title{Aberration-free volumetric high-speed imaging of \textit{in vivo}
retina}

\author{Dierck Hillmann\textsuperscript{1,*}, Hendrik Spahr\textsuperscript{2,3},
Carola Hain\textsuperscript{2}, \\
Helge Sudkamp\textsuperscript{2,3}, Gesa Franke\textsuperscript{2,3},
Clara Pfäffle\textsuperscript{2}, \\
Christian Winter\textsuperscript{1}, and Gereon Hüttmann\textsuperscript{2,3,4}}

\maketitle

\lyxaddress{\textsuperscript{1}Thorlabs GmbH, \\
Maria-Goeppert-Straße 9, 23562 Lübeck, Germany}

\lyxaddress{\textsuperscript{2}Institute of Biomedical Optics Lübeck, \\
Peter-Monnik-Weg 4, 23562 Lübeck, Germany}

\lyxaddress{\textsuperscript{3}Medical Laser Center Lübeck GmbH, \\
Peter-Monnik-Weg 4, 23562 Lübeck, Germany}

\lyxaddress{\textsuperscript{4}Airway Research Center North (ARCN), \\
Member of the German Center for Lung Research (DZL), Germany}

\lyxaddress{\textsuperscript{*}dhillmann@thorlabs.com}

\textsf{\textbf{Research and medicine rely on non-invasive optical
techniques to image living tissue with high resolution in space and
time. But so far a single data acquisition could not provide entirely
diffraction-limited tomographic volumes of rapidly moving or changing
targets, which additionally becomes increasingly difficult in the
presence of aberrations, e.g., when imaging retina}}\textsf{\textbf{\textit{
in vivo}}}\textsf{\textbf{. We show, that a simple interferometric
setup based on parallelized optical coherence tomography acquires
volumetric data with 10 billion voxels per second, exceeding previous
imaging speeds by an order of magnitude. This allows us to computationally
obtain and correct defocus and aberrations resulting in entirely diffraction-limited
volumes. As demonstration, we imaged living human retina with clearly
visible nerve fiber layer, small capillary networks, and photoreceptor
cells, but the technique is also applicable to obtain phase-sensitive
volumes of other scattering structures at unprecedented acquisition
speeds.}}

Fourier-domain optical coherence tomography (FD-OCT) images living
tissue with high resolution \cite{Fercher:95,Chinn:97,Haeusler:98}.
Its most important applications are currently in ophthalmology, where
it provides three-dimensional data of the human retina that are not
attainable with any other imaging method. However, especially at high
numerical apertures (NA), aberrations significantly reduce its resolution
and the focal range restricts the volume depth that is obtained in
a single measurement. Although computational methods have been shown
to remove these limitations \cite{Stadelmaier:00,Fienup:03,Ralston:2007,Adie:12,Adie:12:GuideStar,Adeel:2013},
they have hardly been applicable \textit{in vivo}, as these methods
face two major challenges: First, volumes have to be acquired coherently,
i.e., phases must not be influenced by sample motion, but must only
depend on tissue morphology. And second, the aberrations need to be
determined reliably.

Essentially, FD-OCT acquires coherent volumes. It interferometrically
detects backscattered infrared light at multiple wavelengths and computes
its depth-resolved amplitude and phase at one lateral point (A-scan).
To obtain a three-dimensional volume it usually acquires A-scans for
different lateral positions by confocal scanning. If all A-scans are
measured without sample motion, the volume is phase-stable and coherent,
and the advantage of such data was previously shown: degradation of
the lateral resolution by a limited focal depth was eliminated by
interferometric synthetic aperture microscopy (ISAM) \cite{Ralston:2007,Adeel:2013},
and later, Adie et al.~\cite{Adie:12,Adie:12:GuideStar} corrected
aberrations.

However, a moving sample destroys the phase-stability and makes the
acquired OCT dataset virtually not coherent, which held back \textit{in
vivo} applications of ISAM \cite{Shemonski:14:Stability1,Shemonski:14:Stability2}.
Sample tracking and motion correction improved the phase-stability
\cite{Shemonski:14}, and even the photoreceptor layer of living human
retina was imaged \cite{Shemonski:15}. But to achieve sufficient
phase-stability, imaging had to be limited to a single \textit{en
face} layer of the retina, and data with little tissue motion had
to be selected. No three-dimensional tomography was possible.

To acquire a phase-stable three-dimensional volume of targets such
as the human retina \textit{in vivo} and to numerically correct defocus
and aberrations, a further increase in imaging speed is required.
In principle, full-field swept-source OCT (FF-SS-OCT) \cite{Povazay:06}
can acquire data several orders of magnitude faster than confocal
OCT as it removes the lateral scanning by imaging all positions onto
an area camera and it allows higher radiant flux on the sample without
damaging the tissue. But so far, it showed poor image quality and
the available cameras limited its imaging speed and field of view
\cite{Bonin:10}.

Here we show, that a remarkably simple FF-SS-OCT system obtains truly
coherent three-dimensional tomograms of the living human retina with
high image quality. Its acquisition speed is $38.6\,\mathrm{MHz}$
lateral points (A-scans) per second, which exceeds current clinical
systems by several orders of magnitude and is about one order of magnitude
faster than any other OCT system in research \cite{Klein:13,Kocaoglu:14}.
Its fast and phase-stable acquisition allows a computational optimization
of image quality, similar to \cite{Fienup:00} in synthetic aperture
radar, which removes defocus and aberrations and pinpoints all light
to its three-dimensional scattering location in a volume spanning
multiple Rayleigh lengths in depth. No additional hardware tracks
motion, or determines and corrects aberrations. To show the validity
of this approach, we imaged the living human retina at maximum pupil
diameter (7~mm), and obtained images of the nerve fiber layer, small
vascular structures, and the photoreceptor cells with nearly diffraction-limited
resolution. The presented technique visualizes living and moving tissue,
such as the retina, with higher lateral and temporal resolution than
previously possible. In addition the detected phase of the scattered
light can provide valuable data on small subwavelength changes in
axial direction. Thereby the technique can visualize dynamic \textit{in
vivo} processes that were previously inaccessible.

\section{Data acquisition and processing}

\begin{figure}
\begin{centering}
\includegraphics{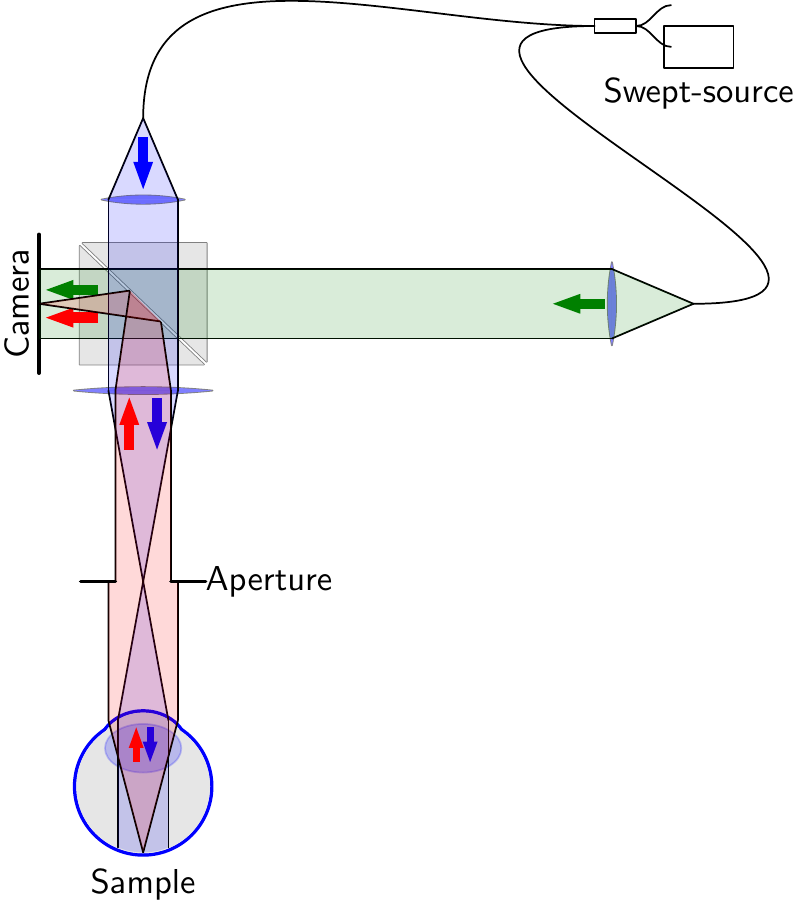}
\par\end{centering}

\caption{\label{fig:Setup}Setup of the full-field swept-source OCT for retinal
imaging. Light from a tunable light source is split into reference
(green) and sample arm (blue); the sample light illuminates the retina
and the backscattered light (red) is imaged onto the camera where
it is superimposed with the reference light. }
\end{figure}
To acquire an entire three-dimensional volume coherently by FF-SS-OCT,
interference images of light backscattered by the retina and reference
light were acquired at multiple wavelengths. To this end, the interference
pattern was generated in a simple Mach-Zehnder type setup (see Methods)
as shown in Fig.~\ref{fig:Setup}. For a single volume, a high-speed
camera (Photron FASTCAM SA-Z) recorded $512$ of these interferograms
during the wavelength sweep of a tunable laser (Superlum Broadsweeper
BS-840-1), covering $50\,\mathrm{nm}$ and centered at $841\,\mathrm{nm}$.
Images were acquired with $896\times368$ pixels at $60,000$ frames
per second, which results in an acquisition rate of 117 volumes per
second and corresponds to $38.6\,\mathrm{MHz}$ A-scan rate and $9.9\,\mathrm{GHz}$
voxel rate.

We first reconstructed the acquired data using standard OCT processing
(see Methods). Most importantly, a Fourier transform of the acquired
data along the wavenumber axis reconstructed the sample volume. Afterwards
computational corrections of axial motion maximized image quality;
this was inevitable, even at these acquisition rates. The resulting
image volumes were coherent and contained the correct phases, but
still suffered from reduced quality due to a limited focal depth and
wavefront aberrations.

\section{Principle of aberrations and their correction}

\begin{figure*}
\begin{centering}
\includegraphics{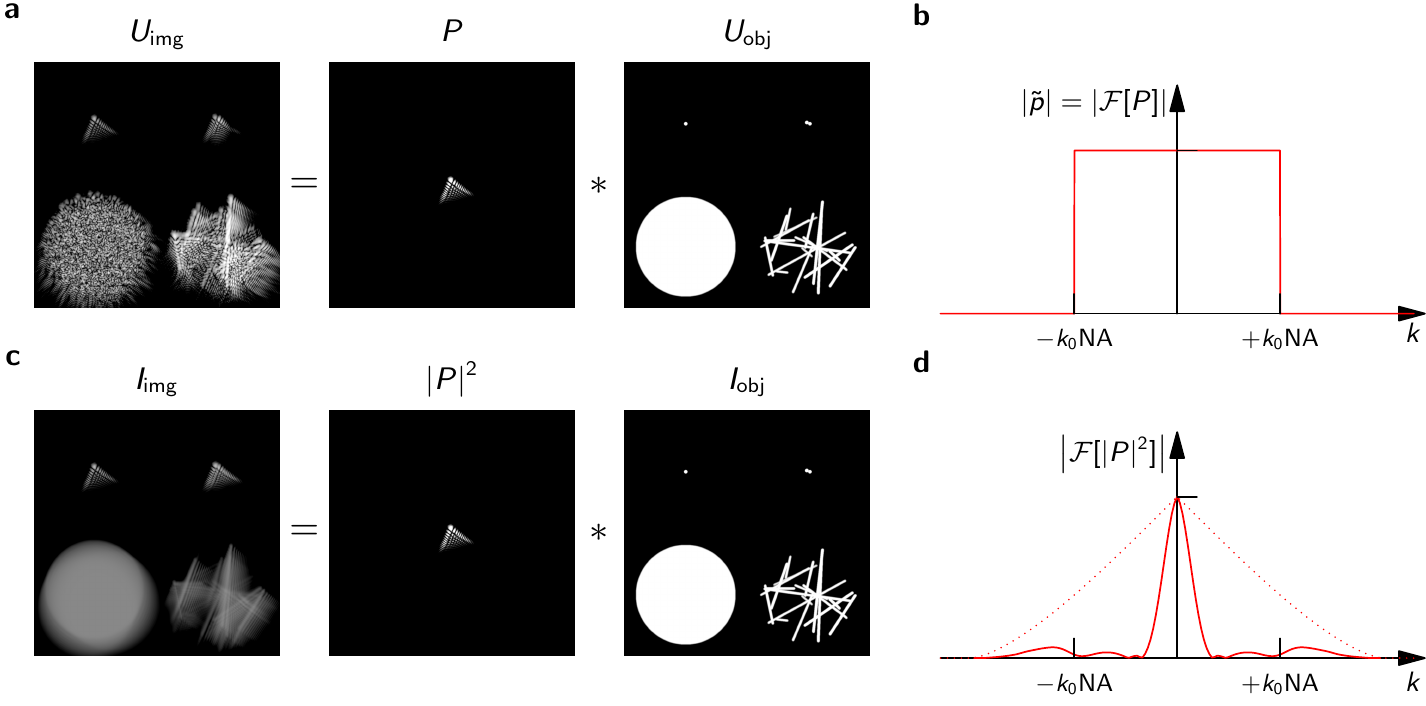}
\par\end{centering}

\caption{\label{fig:Imaging}Illustration of coherent and incoherent imaging
and the corresponding transfer functions. a)~Coherent case: the obtained
image $U_{\mathrm{img}}$ is the convolution of the wave field in
the object plane $U_{\mathrm{obj}}$ with the point spread function
$P$; the circle (bottom-left) has laterally varying random phases.
b)~The Fourier transform of $P$, i.e.~$\tilde{p}$, has unit magnitude
within the aperture; its phase varies and results in aberrations.
c)~Incoherent case: The intensities of the wave in the object plane
$I_{\mathrm{obj}}$ convolved with the intensity of $P$ give the
image $I_{\mathrm{img}}$; no interference effects occur. d)~The
incoherent transfer function is in general complex with varying magnitudes.}
\end{figure*}
Image formation of a single depth layer in a coherent OCT volume is
described by coherent imaging theory (see e.g. \cite{Goodman:05}).
It is assumed that the detected complex wave field in the image plane
$U_{\text{img}}$ is a convolution of the wave field in the object
plane $U_{\text{obj}}$ and the aberrated complex point spread function
(PSF) $P$ 
\begin{equation}
U_{\text{img}}\left(\vec{x}\right)=P\left(\vec{x}\right)*U_{\text{obj}}\left(\vec{x}\right)\mbox{,}\label{eq:ComplexConvolution}
\end{equation}
where $\vec{x}$ is the lateral position in the respective plane when
neglecting magnification. In general a limited aperture or aberrations
will broaden or distort the PSF. The convolution with this PSF not
only degrades the resolution of $U_{\mathrm{img}}$, but also causes
artificial structures by interference of the aberrated image points,
which eventually introduces speckle noise if the object structures
cannot be resolved (Fig.~\ref{fig:Imaging}a).

The effect of aberrations on coherent imaging systems is even better
visible in the frequency domain. If the phase of $U_{\mathrm{img}}\left(\vec{x}\right)$
is recorded correctly, the convolution theorem can be applied to the
coherent imaging equation (\ref{eq:ComplexConvolution}). It then
translates to 
\begin{equation}
\tilde{u}_{\text{img}}\left(\vec{k}\right)=\tilde{p}\left(\vec{k}\right)\cdot\tilde{u}_{\text{obj}}\left(\vec{k}\right)\mbox{,}\label{eq:FFTCoherentImage}
\end{equation}
with $\vec{k}$ being the Fourier conjugate variable of $\vec{x}$,
$\tilde{u}$ indicating the respective Fourier transforms of $U$,
and $\tilde{p}$ being the coherent amplitude transfer function of
the imaging system, i.e., the Fourier transform of $P$. Within the
aperture the coherent amplitude transfer function $\tilde{p}\left(\vec{k}\right)$
only has a phase component, 
\begin{equation}
\tilde{p}\left(\vec{k}\right)=\begin{cases}
\exp\left(\mathrm{i}\phi\left(\vec{k}\right)\right) & \left|\vec{k}\right|\le k_{0}\cdot\mathrm{NA}\\
0 & \mbox{otherwise}
\end{cases}\mbox{, }\label{eq:ATF}
\end{equation}
with $k_{0}$ being the (center) wavenumber of the light and $\phi\left(\vec{k}\right)$
being a suitable phase function; outside the aperture, the amplitude
transfer function is $0$ as no light is transmitted (Fig.~\ref{fig:Imaging}b),
and hence, the multiplication with $\tilde{p}\left(\vec{k}\right)$
in equation (\ref{eq:FFTCoherentImage}) effectively low-pass filters
the image $U_{\mathrm{img}}$. Aberrations of the imaging system including
defocus will only change the phase of $\tilde{p}\left(\vec{k}\right)$,
and consequently its effect on the image is completely reversed by
multiplication of $\tilde{u}_{\mathrm{img}}$ with the conjugate complex
of $\tilde{p}\left(\vec{k}\right)$, which corresponds to a deconvolution
of equation (\ref{eq:ComplexConvolution}). Since the signal energy
at all transmitted spatial frequencies is not attenuated, i.e., the
absolute value of $\tilde{p}\left(\vec{k}\right)$ is not decreased
within the aperture, the reconstruction is lossless, even in the presence
of noise. However, to achieve this $\phi\left(\vec{k}\right)$ needs
to be known.

The corresponding incoherent process illustrates the difference to
a deconvolution in standard image processing. With an incoherent light
source, only the convolution of the scattering intensities $I_{\text{obj}}$
with the squared absolute value of the PSF is detected in the image
$I_{\mathrm{img}}$, 
\[
I_{\text{img}}\left(\vec{x}\right)=\left|P\left(\vec{x}\right)\right|^{2}*I_{\text{obj}}\left(\vec{x}\right)\mbox{.}
\]
During incoherent imaging, defocus and aberrations only cause a loss
of contrast for small structures (Fig.~\ref{fig:Imaging}c), no additional
interference and no speckle noise occur. However, the optical transfer
function, i.e., the Fourier transform of $\left|P\left(\vec{x}\right)\right|^{2}$,
is in general complex, and may contain small or even zero values (Fig.~\ref{fig:Imaging}d).
Hence the effect of aberrations on image quality cannot be inverted
without loosing information or increasing noise. In this context,
it is remarkably that a simple multiplication with the complex conjugate
of $\tilde{p}$ inverts the coherent process, despite of speckle noise.

The theory of coherent imaging also applies to the signal formation
in FD-OCT. Here, $\tilde{p}\left(k\right)$ is a function of the spectral
wavenumber $k$, and the Fourier conjugate to $k$ is the depth. Shape
and width of $\tilde{p}\left(k\right)$ are given by the spectrum
of the light source, which also determines the axial PSF and thus
the resolution. Similar to coherent aberrations, an additional phase
term is introduced if reference and sample arm have a group velocity
dispersion mismatch or, which is relevant for FF-SS-OCT, if the sample
moves axially \cite{Hillmann:12}. As for aberrations, this is corrected
losslessly by multiplication of the spectra with the conjugated phase
term, if it is known.

\section{Aberration detection}

To computationally correct aberrations in coherent imaging, it is
crucial to determine the aberration-related phase function $\phi\left(\vec{k}\right)$
first and various approaches have been developed to determine it.
One approach uses single points in the image data as guide stars \cite{Liu:13,Adie:12:GuideStar},
which is the numerical equivalent to a direct aberration measurement
with a wavefront sensor. Although photoreceptors can be used as guide
stars in not too severely aberrated retinal imaging \cite{Shemonski:15},
a guide star is usually not available in other retinal layers or other
tissue. A second approach cross-correlates low-resolution reconstructions
of the aberrated image from different sub-apertures to estimate the
phase front. It worked fairly well in digital holography \cite{Tippie:11},
in FF-SS-OCT \cite{Kumar:13}, and as a rough first estimation for
\textit{in vivo} photoreceptor imaging \cite{Shemonski:15}, and also
to correct dispersion and axial motion in FF-SS-OCT \cite{Hillmann:12}.
However, these low-resolution images of scattering structures show
usually independent speckle patterns, which carry no information on
the aberrations and which limit the precision of the phase front determination.
Additionally, the uncertainty relation couples spatial resolution
and accuracy of the resulting $\phi\left(\vec{k}\right)$; increasing
resolution decreases accuracy of the phase $\phi\left(\vec{k}\right)$,
and vice versa. 

Here, we iteratively optimized the image quality to obtain the correcting
phase, which provided very good results. Although it is computationally
expensive, this idea was already applied to Digital Holography \cite{Thurman:08},
synthetic aperture radar (SAR, \cite{Flores:91,Fienup:00}), and also
scanning OCT to correct aberrations \cite{Adie:12} and dispersion
mismatch \cite{Wojtkowski:04}. In this approach, a wavefront $\phi\left(\vec{k}\right)$
is assumed, and equation (\ref{eq:FFTCoherentImage}) is inverted
by multiplying $\tilde{u}_{\mathrm{img}}$ with $\tilde{p}^{*}\left(\vec{k}\right)=\exp\left(-\mathrm{i}\phi\vec{\left(k\right)}\right)$.
After an inverse Fourier transform a corrected image is obtained,
which can be evaluated for image quality. The task is to find the
$\phi\left(\vec{k}\right)$ that gives the best quality and thus corrects
aberrations. 

For this approach, a metric $S\left[U_{\mathrm{img}}\left(\vec{x}\right)\right]$
describing image-sharpness, a parameterization of the phase error
$\phi\left(\vec{k}\right)$, and finally, an optimization algorithm
are required, and their choice influences quality of the results and
performance of the approach. The metric $S$ needs to be minimal (or
maximal) for the aberration-free and focused image, even in the presence
of speckle noise. A parametrization of the phase front $\phi\left(\vec{k}\right)$
keeps the dimensionality of the problem low and thus prevents over-fitting;
still, it needs to describe all relevant aberrations. Finally, a robust
optimization technique must find the global minimum of the metric
without getting stuck in local minima. As the number of free parameters
increases with higher aberration order, the global optimization becomes
more difficult and increasingly time consuming; the algorithm performance
is therefore crucial.

A variety of metrics and image-sharpness criteria have been used in
previous research for coherent and incoherent imaging \cite{Muller:74,Paxman:88,Flores:91,Fienup:00,Fienup:03}.
For a normalized complex image given by $U_{mn}$ at pixel $m,n$,
a special class of metrics \cite{Fienup:03} only depends on the sum
of transformed image intensities (see also Methods):

\begin{equation}
S\left[U_{mn}\right]=\sum_{m,n}\Gamma\left(\left|U_{mn}\right|^{2}\right)\mbox{.}\label{eq:Metric}
\end{equation}
Here, we used the Shannon entropy given by $\Gamma\left(I\right)=-I\log I$
\cite{Flores:91}, although we observed similar performance when choosing
$\Gamma\left(I\right)=I^{\gamma}$ with a $\gamma<1$ \cite{Muller:74}.
When these metrics were minimal, we observed good image quality, despite
of speckle dominated data.

As parametrization, the phase function $\phi\left(\vec{k}\right)$
was expressed in Zernike polynomials. These are established in the
description of optical aberrations including defocus, and their use
gave good results and performance during optimization. Zernike polynomials
up to \nth{8} radial degree were used, excluding piston, tip, and
tilt, which results in $42$ degrees of freedom.

The optimization has to find the Zernike coefficients describing $\phi\left(\vec{k}\right)$
that give the absolute minimum of the metric $S$ for the acquired
data. To achieve this, we used a two-step approach. At first a simplex-downhill
algorithm was applied \cite{Nelder:65}. This algorithm follows the
global trend of the metric function and is thereby insensitive to
local minima. Once being close to the global minimum, a gradient-based
algorithm was used, which significantly boosted performance. Here,
we used the conjugate gradient method \cite{Hestenes:52}. A useful
property of metrics described by equation (\ref{eq:Metric}) is that
their complete gradient with respect to the Zernike coefficients can
be computed efficiently; it requires only a single additional Fourier
transform (see \cite{Fienup:00,Fienup:03} and Methods).

If aberrations were too strong and the degrees of freedom too large
the optimal phase front could not be determined in this way. Instead
the optimization was first performed with a computationally reduced
numerical aperture, which gave robust results with reduced lateral
resolution. Afterwards, the optimization was repeated several times
while increasing the NA. This way even large aberrations including
all 42 degrees of freedom could be corrected. 

In general the assumption of a laterally invariant PSF is not valid,
and equation (\ref{eq:ComplexConvolution}) only holds for small volumes.
The entire data were therefore divided into sufficiently small regions
(see Methods), which were then corrected independently. By stitching
these, aberration-free data for the complete recorded volume were
obtained.

\section{Results}

\begin{figure}
\begin{centering}
\includegraphics{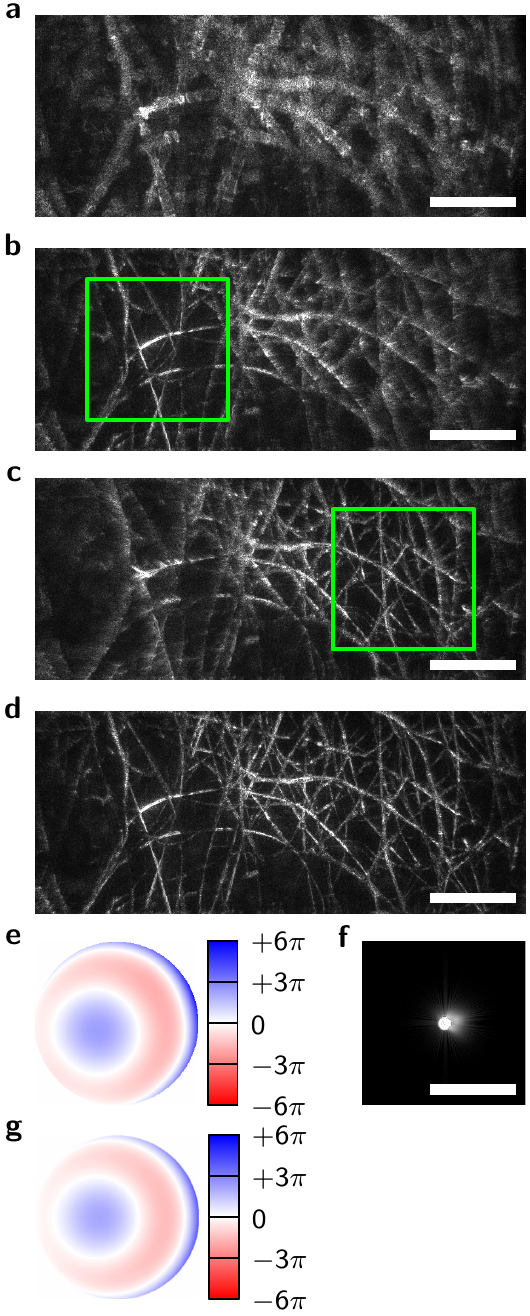}
\par\end{centering}

\caption{\label{fig:LensTissue}\textit{En face} full-field OCT images from
a recorded volume of lens tissue before and after aberration removal.
a)~Before aberration correction. b)-c)~After correcting aberrations
for different regions (green squares). d)~After stitching multiple
corrected regions. e)~Wavefront as determined for (c) without defocus.
f)~Corresponding PSF. g)~Simulation of the expected wavefront using
numerical raytracing through the objective lens. Scale bars are $0.5\,\mathrm{mm}$. }
\end{figure}
To demonstrate the accuracy of the algorithms, we first imaged lens
cleaning tissue with a simple achromatic lens at an NA of 0.15, which
introduced significant spherical aberrations (Fig.~\ref{fig:LensTissue}).
Before correction \textit{en face} images were severely blurred (Fig.~\ref{fig:LensTissue}a),
but the optimization restored the fiber structures of the lens tissue
in a certain field of view. Since the aberrations were not translation
invariant, different sub-images were corrected independently (Fig.~\ref{fig:LensTissue}b
and c), and by stitching these the entire image field was obtained
(Fig.~\ref{fig:LensTissue}d). 

Wavefront and PSF resulting from the aberration determination of Fig.~\ref{fig:LensTissue}c
are shown in Figs.~\ref{fig:LensTissue}e and f, respectively. The
obtained wavefront was compared to a raytracing simulation (Zemax,
Fig.~\ref{fig:LensTissue}g) with almost identical results. A slight
lateral misalignment of the imaging optics explains remaining differences. 

\begin{figure}
\begin{centering}
\includegraphics{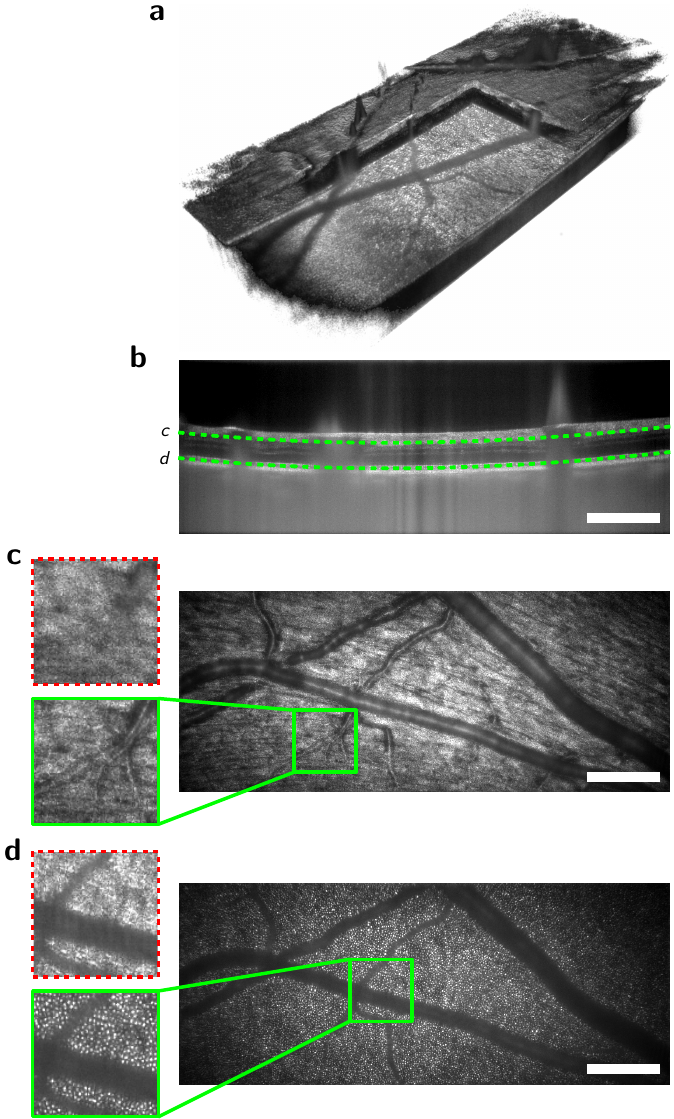}
\par\end{centering}

\caption{\label{fig:RetinaLowNA}Retina volume acquired by FF-SS-OCT at NA
0.1 after aberration removal. a)~Rendered volume.~b)~B-scan (sectional
view) from the volume; the dashed green lines indicate the location
of the \textit{en face} images shown in c) and d). c)-d)~Aberration-corrected
\textit{en face} images and magnification of a small area in the green
boxes; the same area is shown before aberration removal (dashed red
boxes); nerve fiber layer (c) and photoreceptor mosaic (d) are clearly
visible. Scale bars are 0.5~mm.}
\end{figure}
\begin{figure*}
\begin{centering}
\includegraphics{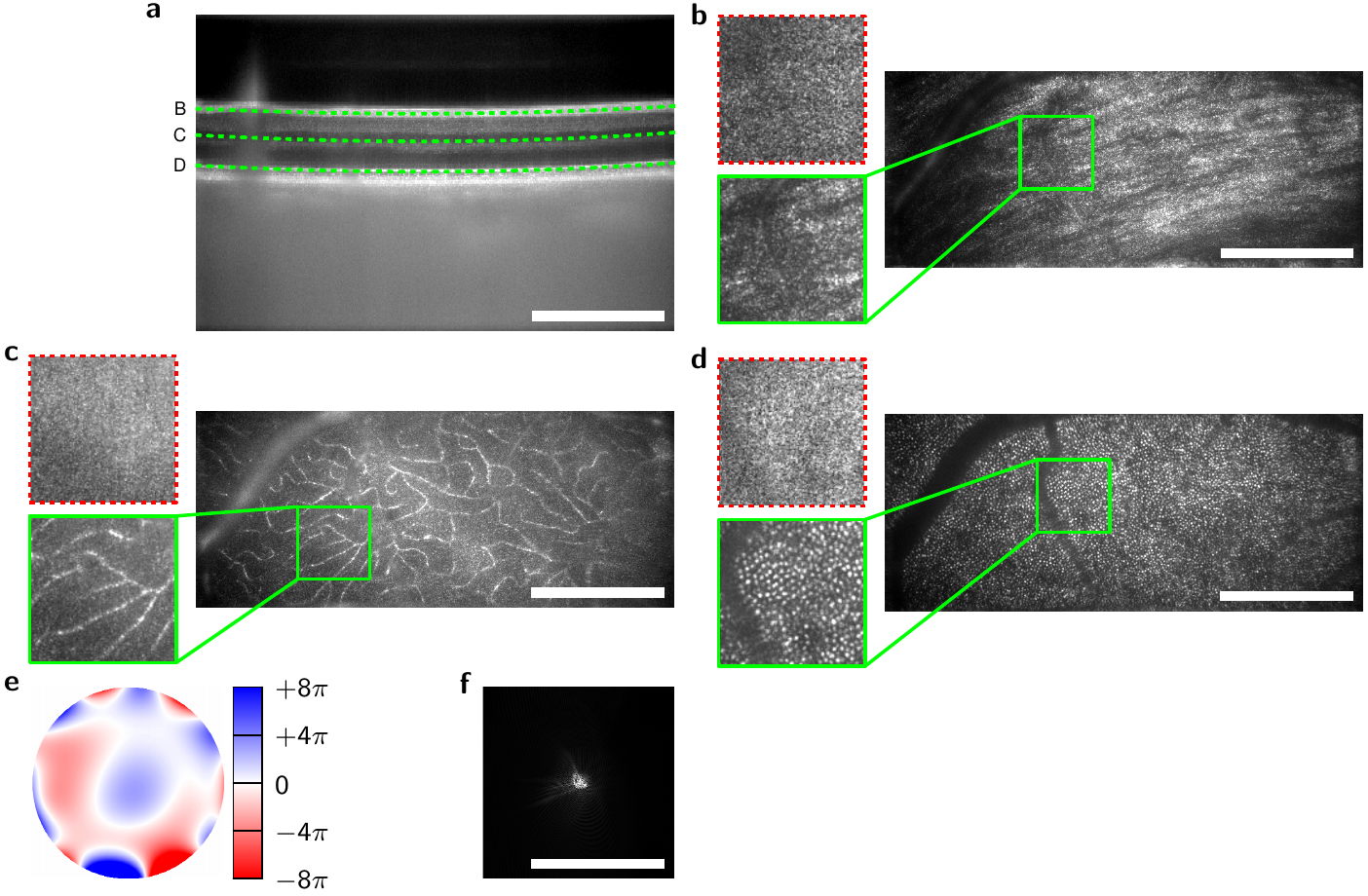}
\par\end{centering}

\caption{\label{fig:RetinaHighNA}Retinal volume acquired by FF-SS-OCT at NA
0.2 corresponding to $7\,\mathrm{mm}$ pupil diameter after aberration
removal. a)~B-scan (sectional view) from the recorded volume; the
dashed green lines indicate the location of the \textit{en face} images
shown in (b)-(d). b)-d)~Aberration-corrected \textit{En face} images
with magnification of a small area in the green boxes; the same area
is shown before aberration removal (dashed red boxes); nerve fiber
layer (B), small capillaries (c) and photoreceptor mosaic (d) are
only visible after correction. e)~Wavefront used for aberration correction
of the green box in (d) with defocus removed. f)~Corresponding PSF.
Scale bars are $0.5\,\mathrm{mm}$.}
\end{figure*}
We then imaged the retina of a young healthy volunteer \textit{in
vivo}. Two datasets were acquired, first at $14^{\circ}$ periphery
and NA 0.1 (Fig.~\ref{fig:RetinaLowNA}), and later at $8^{\circ}$
and 0.2 NA corresponding to $7\,\mathrm{mm}$ diameter of the maximally
dilated pupil (Fig.~\ref{fig:RetinaHighNA}). Without the aberration
correction the volumes are laterally blurred in all layers of the
retina with image degradation being significantly stronger in the
high NA data, where hardly any lateral structures were visible at
first. Several of the layers were aberration corrected, including
the nerve fiber layer (NFL, Figs.~\ref{fig:RetinaLowNA}c, \ref{fig:RetinaHighNA}b),
small capillaries (Fig.~\ref{fig:RetinaHighNA}c), and the photoreceptor
layer (Figs.~\ref{fig:RetinaLowNA}d, \ref{fig:RetinaHighNA}d).
The optimization algorithm improved image quality nearly to diffraction
limit, showing otherwise invisible structures. In particular, the
structure of the nerve fiber layer and small capillaries became visible
and single photoreceptor cells were identified.

However, the sectional images (B-scans) in Figs.~\ref{fig:RetinaLowNA}b
and \ref{fig:RetinaHighNA}a also show a disadvantage of the FF-SS-OCT
technology. In the choroid of the imaged retina, barely any structures
are visible, which is caused by multiple scattered photons \cite{Karamata_I:2005,Karamata_II:2005}.
In addition, artifacts in the larger vessels, caused by the Doppler
effect, decrease overall image quality.

The applied method for determination of the phase function $\phi\left(\vec{k}\right)$
was robust on scattering structures and gave good results correcting
the aberrations induced by the eye. The computation time with the
current implementation depended on the size of the region taken into
account. For a small area of a single volume the computation time
ranged from a few seconds to a less than a minute on a standard desktop
CPU but can certainly be further reduced.

\section{Discussion and conclusion}

For the first time, completely phase-stable volumetric data of human
retina were acquired \textit{in vivo}. This was possible by a currently
unmatched A-scan rate reaching $38.6\,\mathrm{MHz}$, corresponding
to $117$ volumes per second with $84$ million voxels per volume.
This phase-stable data allowed us to correct aberrations and to remove
the effects of a limited focal depth. The demonstrated image quality
optimization worked in retinal imaging to correct the nerve fiber
layer, small capillaries and the photoreceptor cells and is largely
independent of the imaged structure. Only specular reflections not
filling the aperture, or the complete absence of any signal disturbs
the correction. This way, in only a single measurement all retinal
layers were obtained with diffraction limited resolution. 

Currently, the pixel number of the camera that can be used at the
required acquisition rate, and the tuning range of the laser limit
the system capabilities. The former restricts the field of view, the
latter results in a low axial resolution of approximately $10\,\mathrm{\mu m}$
in tissue, especially when compared to the diffraction-limited lateral
resolution of $2.6\,\mathrm{\mu m}$ at 0.2 NA. However, we expect
that both limitations will be addressed by future technological advances. 

Compared to scanning systems, disadvantages of FF-SS-OCT are Doppler
artifacts and a higher sensitivity to multiple scattered photons,
although the latter was not as severe as we anticipated. On the other
hand, system complexity is reduced significantly and no moving parts
are involved. Our setup profits from a recently available high-speed
CMOS camera, which is its most advanced and expensive component. As
its technology matures, availability might increase making FF-SS-OCT
a cost-effective alternative to complex OCT systems that implement
adaptive optics (AO-OCT. e.g.~\cite{Kocaoglu:14}) with deformable
mirrors and wavefront sensors. Partially due to this complexity, AO-OCT
has not yet found wide spread use in neither clinics nor clinical
research. 

However, fully coherent high-speed tomography not only visualizes
dynamic processes with diffraction limited resolution, but will also
provide new contrast mechanisms that rely on fast and small changes
of scattering properties or of optical pathlengths. Hence FF-SS-OCT
can contribute to numerous areas, e.g., measure tissue responses to
photocoagulation \cite{Mueller:12}, detect heart-beat-induced pressure
waves in order to probe vascular status \cite{Kotliar:2013-5943,Spahr:15},
or obtain data for optophysiology \cite{Bizheva:2006-3693}.

\section{Methods}

\paragraph*{{\small{}Setup}}

{\small{}A simple Mach-Zehnder interferometer setup was used for full-field
OCT imaging of the retina (Fig.~\ref{fig:Setup}). Light from a tunable
light source was split by a fiber coupler into reference and sample
arm. The light in the reference arm was collimated and brought onto
the camera through a beam splitter cube. Light from the sample arm
illuminated the retina through the same beam splitter cube and the
imaging optics with a parallel beam. The backscattered light from
the retina was imaged onto the camera with numerical apertures (NA)
ranging from $0.1$ to $0.2$, the latter corresponding to the maximum
aperture of the human eye ($\sim7\,\mathrm{mm}$) which was only achieved
during mydriasis. The illuminated field on the retina was imaged onto
the area-of-interest of the camera, maximizing light efficiency. Polarization
of both arms was matched to enhance interference contrast and sensitivity.}{\small \par}

{\small{}The light source (Superlum BroadSweeper BS-840) was tunable
over $50\,\mathrm{nm}$ with a central wavelength of $840\,\mathrm{nm}$,
which resulted in about $10\,\mathrm{\mu m}$ axial resolution in
tissue. In combination with a semiconductor amplifier approximately
$20\,\mathrm{mW}$ were coupled into the interferometer, illuminating
the extended area on the retina with approximately $5\,\mathrm{mW}$.
The high speed CMOS camera (Photron FASTCAM SA-Z) achieved a frame
rate of $60,000\,\mathrm{fps}$ at a resolution of $896\times368\,\mathrm{pixels}$.
For each volume $512$ images were acquired, each at a different wavenumber
in the tuning range of the light source. The acquisition speed is
thus equivalent to $38.6\,\mathrm{MHz}$ A-scan rate. For each measurement
$50$ volumes were imaged.}{\small \par}

\paragraph{{\small{}Reconstruction}}

{\small{}At first a coherent average of the 50 acquired volumes was
computed and the resulting mean was subtracted from all volumes. This
removed fixed and phase stable artifacts in the images, while leaving
the signals of the moving retina intact. Afterwards, in analogy to
FD-OCT signal processing, the OCT volumes were reconstructed by Fourier
transforming the acquired $512$ images along the wavenumber axis
giving the depth information at each pixel of the image. }{\small \par}

{\small{}The data was then corrected for group velocity dispersion
mismatch in reference and sample arm and slight axial bulk motion.
This was done using the same optimization approach that was used for
aberration correction by approximating the dispersion phase function
by a polynomial of order 16. The resulting volumes were axially and
laterally shifted to maximize the correlation of their absolute values
by using the Fourier shift theorem. This ensured that the layers of
the retina were at identical depth positions and made selection of
the input regions for the aberration correction easier. Finally, after
aberration correction, this correlation maximization was repeated
giving more precise results due to the smaller lateral structures.
This ensured that structures were in the same place so that absolute
values of the data could be averaged. Finally, 2 to 6 layers of the
respective retinal structures were averaged giving the presented }\textit{\small{}en
face}{\small{} images.}{\small \par}

\paragraph{{\small{}Aberration Correction}}

{\small{}Complex }\textit{\small{}en face}{\small{} images, i.e.~slices
at a certain depth given by $U_{mn}$ at pixel $m,n$, were taken
as input for the aberration removal step. In general multiple layers
were used to improve the overall signal of the metric. Small sub-volumes
of about $96\times96$ pixels with 6 to 10 layers turned out to be
sufficient to correct aberrations for the low NA retinal image. For
the high NA retinal images sub-volume size was about $128\times128$
pixels with 4 to 10 layers. At first all volumes were corrected individually
choosing a central region to determine the aberrations up to \nth{8}
radial degree (42 degrees of freedom). Afterwards the entire data
were stitched together of small sub-volumes, each corrected individually.
For this last step only aberrations to \nth{5} radial degree were
taken into account and the metric was averaged over all 50 volumes.}{\small \par}

{\small{}The aberration correction itself is similar to a process
that was previously shown for SAR imaging by Fienup \cite{Fienup:00,Fienup:03}.
For a given phase front $\phi_{\mu\nu}$, where $\mu,\nu$ describe
pixels in lateral frequency space, the phase front was evaluated by
\[
\phi_{\mu\nu}=\sum_{k}a_{k}Z_{\mu\nu}^{(k)}\mbox{,}
\]
with $Z_{\mu\nu}^{(k)}$ being the Zernike polynomials, $a_{k}$ being
coefficients describing the phase front, and $k$ enumerating the
different Zernike polynomials. The image was then reconstructed by
\[
U_{mn}=\sum_{\mu,\nu}F_{m\mu}^{-1}F_{n\nu}^{-1}\left[\tilde{u}_{\mu\nu}\mathrm{e}^{\mathrm{i}\phi_{\mu\nu}}\right]\mbox{,}
\]
where $\tilde{u}_{\mu\nu}$ denotes the two-dimensional discrete Fourier
transform of $U_{mn}$ and $F_{m\mu}=\exp\left(\mathrm{i}m\mu/\left(2\pi N\right)\right)$
is the discrete Fourier matrix. The sum can be performed by a fast
Fourier transform (FFT). }{\small \par}

{\small{}The image quality or sharpness of the }\textit{\small{}en
face}{\small{} images was evaluated with the help of the normalized
image intensity $I_{mn}=\left|\nicefrac{U_{mn}}{\sum_{m,n}\left|U_{mn}\right|}\right|^{2}$
by the Shannon entropy $S$ given by
\begin{multline}
S\left[I_{mn}\right]=\sum_{m,n}\Gamma\left(I_{mn}\right)=-\sum_{m,n}I_{mn}\log I_{mn}\mbox{,}\\
\mbox{with }\Gamma\left(I\right)=-I\log I\mbox{,}\label{eq:ShannonEntropy}
\end{multline}
which is supposed to be minimized for optimal imaging quality \cite{Flores:91}.
In analogy to the demonstration by Fienup \cite{Fienup:00}, the gradient
could be efficiently evaluated by 
\begin{multline*}
\frac{\partial S\left[I_{mn}\right]}{\partial a_{k}}=\\
-2\sum_{\mu,\nu}\Im\Biggl[\sum_{m,n}\left.\frac{\partial\Gamma\left(I\right)}{\partial I}\right|_{I=I_{mn}}U_{mn}^{*}F_{m\mu}^{-1}F_{n\nu}^{-1}\tilde{u}_{\mu\nu}\Biggr]Z_{\mu\nu}^{(k)}\mbox{.}
\end{multline*}
}{\small \par}

\section{Acknowledgements}

This research was sponsored by the German Federal Ministry of Education
and Research (Innovative Imaging \& Intervention in early AMD, contract
numbers 98729873C and 98729873E).

\section{Author contributions}

D.H.~contributed theoretical and mathematical basics, worked on the
optical setup, wrote the manuscript, analyzed the data, and helped
to obtain funding. H.Sp.~and G.F.~worked on the optical setup, collected
data and reviewed the manuscript. C.H., H.Su., and C.P.~worked on
the optical setup and collected data. C.W.~performed Zemax simulations.
G.H.~contributed to the theoretical basics, partly wrote, reviewed
and edited the manuscript, as well as obtained funding.

\section{Competing financial interests}

D.H. and C.W. are working for Thorlabs GmbH, which produces and sells
OCT systems. D.H. and G.H.~are listed as inventors on a related patent
application (application no. PCT/EP2012/001639). 

\bibliographystyle{naturemag}
\bibliography{Aberrations_arXiv}

\end{document}